\documentclass[%
 reprint,
 amsmath,amssymb,
 aps,
nofootinbib,
floatfix,
]{revtex4-1}

\usepackage{graphicx}% Include figure files
\usepackage{dcolumn}% Align table columns on decimal point
\usepackage{bm}% bold math
\usepackage{scalerel}

\newcommand{\boldu}{{\bf u}}

\newcommand{\boldT}{{\bf T}}
\newcommand{\boldv}{{\bf v}}

\newcommand{\boldD}{{\bf D}}
\newcommand{\boldE}{{\bf E}}

\begin{document}
%\preprint{APS/123-QED}
\title{{\vspace*{-8mm}}The generation and sustenance of electric fields in 
sandstorms}% Force line breaks with \\
\author{{\vspace*{-4mm}}Mustafa Mutiur Rahman}
%\email{Mustafa.Rahman@kaust.edu.sa}
\author{Wan Cheng}
\altaffiliation{ {\vspace*{-.7cm} \\} Also affiliated with Graduate Aerospace Laboratories, California Institute of Technology, CA, 91125, USA}
\author{Ravi Samtaney}
\email{Ravi.Samtaney@kaust.edu.sa}
 \affiliation{%
Division of Physical Science \& Engineering,\\ King Abdullah University of Science \& Technology, Saudi Arabia 23955-6900
}%

\date{\today}% It is always \today, today,
\begin{abstract}
{\vspace*{-4mm}}
Sandstorms are frequently accompanied by the generation of intense electric fields  and lightning. In a very narrow region close to the ground level, sand particles undergo a charge exchange mechanism whereby larger (resp. smaller) sized sand grains become positively (resp. negatively) charged are then entrained by the turbulent fluid motion. Our central hypothesis is that differently sized sand particles get differentially transported by the turbulent flow resulting in a large-scale charge separation, and hence a large-scale electric field. We utilize our simulation framework, comprising of large-eddy simulation of the turbulent atmospheric boundary layer along with sand particle transport and an electrostatic Poisson solver, to investigate the physics of electric fields in sandstorms and thus, to confirm our hypothesis. We utilize the simulation framework to investigate electric fields in weak to strong sandstorms that are characterized by the number density of the sand particles. Our simulations reproduce observational measurements of both mean and RMS fluctuation values of the electric field. We propose a scaling law in which the electric field scales as the two-thirds power of the number density that holds for weak-to-medium sandstorms. 
\end{abstract}

\pacs{Valid PACS appear here}% PACS, the Physics and Astronomy
\maketitle

\begin{figure}[htb!]
\centering
%\vspace*{-1mm}
%\begin{tabular}{c}
%{\includegraphics[width=1.\linewidth]{figures/{0HeightVsMean_mod_eVsSqrRtMSqr_mod_eVs_mod_e9_18}.pdf}} \\ 
{\includegraphics[width=1.\linewidth]{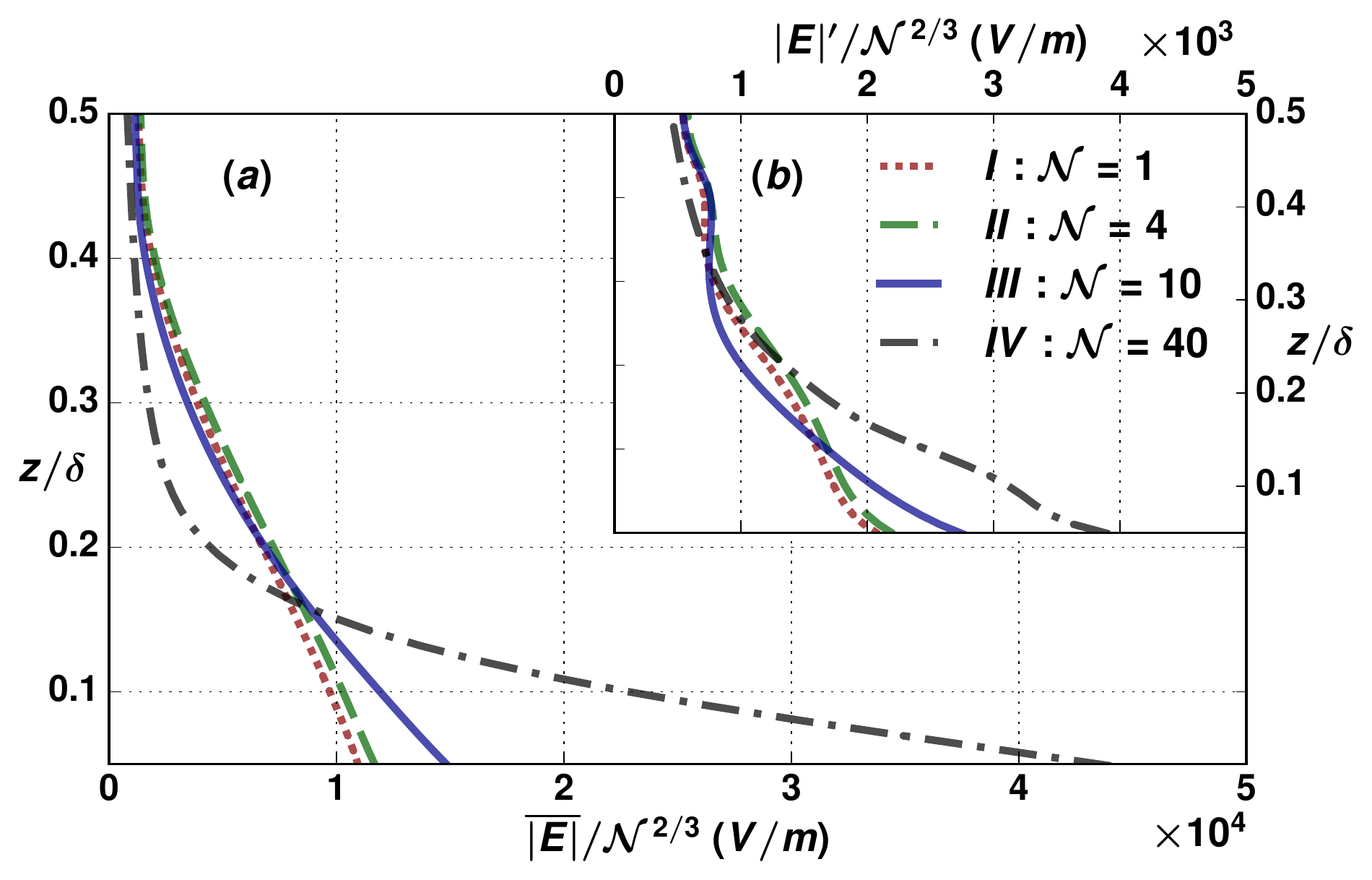}} \\ 
\vspace*{-4.5mm}
 \caption{Altitude variation of the (compensated) net electric field magnitudes of (a) mean $\overline{|{\boldE}|}$ and (b) RMS fluctuations $|{\boldE}|^\prime$
 scaled by  ${\mathcal{N}}^{\frac{2}{3}}$.
% , where 
% Altitude variation  of scaled (compensated) magnitude of the electric field~($|\Vec{E}|$) scaled by ${\mathcal{N}}^{\frac{2}{3}}$ where ${\mathcal{N}} =N_w/N_1$
% (a) mean and (b) RMS fluctuations.
} % (d) Spanwise Electric Field (e)Streamwise Electric Field
    \label{fig:AltitudeVariationOfScalednetelectricfieldmagnitudesformeanRMSfluctuationsandratios}
    \vspace*{-1.07cm}
    %\twocolumngrid
  \end{figure}

%{\vspace*{-.9cm} \\}
{ \onecolumngrid
{\vspace*{-10mm} }
\twocolumngrid
}
{\it{Introduction.-}}
\label{sec:introduction}
As far back as 1850, Faraday noted in a letter that electric fields
accompanied sandstorms: \textit{``I have received your letter respecting dust storms ...
The quantity of electricity which you obtain is enormous.... That it\,[electricity] accompanies
them\,[dust storms], there is not doubt of; but then, that may be as much in the way of effect
as cause''}%\,(italics by Faraday)
\,\cite{baddeley1860whirlwinds}. 
 He was commenting on the misconception
 that electricity was the cause of
 sandstorms.  
It has since been widely accepted that intense electric fields are generated within sandstorms. 
Zhang {\it et al.}\,\cite{zhang2004experimental} reported 
 maximum average
intensity of electric fields of about $200kV\!\!/m$ with instantaneous values
exceeding $2.5MV\!\!/m$. 
These intense electric fields 
cause adverse effects %impacts
such as wild fires%\,\cite{keith1944classification}
, communication disruption, and even explosions\,\cite{keith1944classification,eden1973electrical}. %\,\cite{keith1944classification,eden1973electrical}.  
% It has also been reported that dangerous shocks arise from touching wire fences and metal objects\,\cite{keith1944classification}.
%The so-called ``dust-devils"
The sandstorms on Mars cause problems for rovers and satellites because of the electric fields\,\cite{jackson2006electrostatic}. %\cite{ruf2009emission}. 
 A sandstorm is a complex meteorological phenomenon that generally involves a storm front, high Reynolds number turbulent flow, the transport of sand, and an accompanying electric fields. 
However, a satisfactory explanation of large-scale electric fields in sandstorms is still lacking. 
Recent theoretical work and small-scale laboratory
experiments\,\cite{pahtz2010particle} have produced simple predictive models for the 
charging of granular materials in collisional flows. However, these small scale laboratory
experiments were conducted in the presence of an external electric field,
and, therefore, cannot predict or explain why large-scale charge
separation and self-sustaining electric fields occur in a sandstorm.
Furthermore, as we explain later, the main body of a 
sandstorms has negligible inter-particle collisions. 
The electric field must exceed the dielectric strength\,($3MV\!/m\!$) of the air to produce a flash of lightning. % in a sandstorm. 
Electric fields of 
such intensity 
 cannot arise from the 
triboelectrification or other charge generating mechanisms 
, as have been proposed in the scientific literature~\cite{kanagy1994electrical}. 
 %
%so far 
%proposed in the scientific literature\,\cite{kanagy1994electrical}. 
%
%{\it{Objective.-} }
%\label{sec:Objective}
An important missing piece of physics is
turbulence. 
We believe that the turbulent transport of differently sized sand particles within a sandstorm is the
key 
 process that generates intense electric fields, 
%Our main hypothesis is that the % following:
%\textit
{where the large scale charge separation of charged sand particles provide the
 necessary charge to sustain large electrical fields.}
To test
this hypothesis, we developed
a simulation solver 
 to 
solve the equations and
carry out large-scale 
  turbulent atmospheric  flow  simulations  , along with the  
transport of sand particles 
%and 
 to estimate the
levels of electric fields.  
% Smaller particle class are negatively charged which are easily carried with the flow and vice-versa for larger size.
  The main objective of this work is to model the electric fields in high Reynolds number atmospheric flows, and it is %we do
   not necessary to account for sandstorm fronts. 
% This is due to the 
% need for detailed simulations, which can substitute expensive and sometimes
%  such impractical experiments.
\\
{\it{Physical characteristics of sandstorms and modeling approach:} }
%the flow in
\label{sec:PhysicsFlowSandstorms}
Sandstorms are turbulent air motions with suspended solid sand particles. The sand particles are entrained by large-scale swirling motions within the turbulent atmospheric boundary layer with a characteristic height\,(denoted by $\delta$)  of O(100$m$). The large scale integral length of a sandstorm is of O($km$).
% Resolving all the turbulent scales of motion, and the effects of solid particles on the turbulent scales of motion from a first principles approach i.e. direct numerical simulation (DNS) is so prohibitively expensive that DNS is simply not a viable choice. 
% The standard atmospheric conditions with free stream wind speeds\,(${U}_{\infty}\!$) 
%  varying from $7m\!/s$ to $25m\!/s$ corresponds to Kolmogorov length scale of $\eta\!\sim\!0.1\!-\!1mm$ setting the smallest dissipative scale. 
 Thus, a typical atmospheric boundary layer Reynolds number %\,($Re_{\delta}$)
  can exceed $10^9$. This is large enough that resolving all the turbulent scales with a direct numerical simulation technique is not practically viable, and hence we resort to computing the fluid turbulence with the large-eddy simulation\,(LES) approach.
Wind tunnel experiments have shown that sand particles smaller than 250$\mu m$ acquire a negative charge, whereas  particles larger than 500$\mu m$ acquire a positive charge\,\cite{zheng2003laboratory}.
Most numerical simulations%\cite{zheng2013electrification,zheng2004theoretical,kok2008electrostatics,bo2013theoretical} 
\,\cite{kok2008electrostatics} related to sandstorms focus on the  saltation mode and creeping mode, but do not consider suspension mode. 
In the suspension mode, sand particles are carried by the air flow and do not settle back to the ground.
Here, we consider the sandstorm as a mixture of solid sand particles in the suspension mode within an atmospheric boundary layer and in a statistical steady state. 
%The vertical fluid velocity is non-zero at the lower boundary of suspension, and any sand particles here will be entrained into the flow. 
Collisions between sand particles mostly occur near the ground in an extremely thin layer\,(O($10-100mm$)) where the volume fraction of sand is large, and within which sand particles exchange electrical charge\cite{zhang2010sand}.
% The collision rate between sand particles %decay 
% diminishes extremely rapidly beyond this height, such that in the suspension mode the sand is essentially a collisionless medium. 
  The collision rate between sand particles diminishes extremely rapidly beyond this height, so that in the suspension mode the sand is essentially suspended as a collisionless medium. 
In the LES approach,
 the smallest resolved eddy\,($\xi$)
 is governed by the discrete size of the computational mesh employed. 
%  As such, the timescale ratio\,($\tau_s/\tau_{\xi}$) is particle timescale\,($\tau_s$) to the 
%  timescale of the smallest resolved eddy scale\,($\tau_{\xi}$) where\,$\tau_{\xi}=\tau_{\eta}(\xi/\eta)^{2/3}$. 
 We determine that the particle-size distribution lies in the Eulerian and Equilibrium-Eulerian range (see classification in Balachandar \& Eaton\,\cite{balachandar2010turbulent}) and the Stokes numbers are small. 
%~$\left(\tau_{\xi}=\tau_{\eta}(\xi/\eta)^{2/3}=(1/Re^{1/2})(\delta/U_{\infty})(\xi/\eta)^{2/3}\right)$.
%Following the approach by Balachandar\& Eaton~\cite{balachandar2010turbulent} and Park {\it et al.}~\cite{park2017simple} we plot, in Figure~\ref{fig:StokesNoAnalysis}, the Stokes number as a function of the ratio of particle size to the Kolmogorov scale. 
%  \begin{figure}[htb!]
%%\vspace*{-4.2mm}
%%{\includegraphics[width=1.\linewidth]{figures/LESStokesNoVsDiamByKolmogorovScale31.pdf}} 
%{\includegraphics[width=1.\linewidth]{figures/Fig1.pdf}} 
%%\vspace*{-6mm}
% \caption{Stokes number as a function of the particle size normalized by Kolmogorov scale. Different curves correspond
% to different simulation conditions ranging from
%    low Reynolds number and high-resolution conditions~($\xi/\eta = 10000$) to high Reynolds number and 
%    low resolution~($\xi/\eta = 720000$). Symbols ($\triangleright$) correspond to particle sizes in LES cases considered.
%}
%%\vspace*{-4mm}
%    \label{fig:StokesNoAnalysis} 
%    \end{figure}
%It is evident that most of the particles size distribution lies in the Eulerian and Equilibrium-Eulerian range, and the Stokes numbers are small. 
We adopt the Eulerian description  for the particulate phase of sandstorms. \\
%Each material (in our case air and sand) is considered as a continuum occupying the same region in space.\\
{\it{Governing Equations.-}} 
%The equations governing the fluid phase of a sandstorm are the filtered incompressible Navier-Stokes equations as follows, 
 The filtered incompressible Navier-Stokes equations describes the fluid phase of a sandstorm as,
%The filtered Navier-Stokes equations as follows. 
%The filtered equations are non-dimensionalized with free stream velocity (${U}_{\infty}$) and boundary layer thickness ($\delta$) as
\vspace*{-0.25cm}
\begin{eqnarray}
 \label{eq:lesChannelfluidmasseqn}
 \!\!\!\! \!\!\! \nabla \! \cdot\! \tilde{\boldu}\!=\!0,\!\partial_t \tilde{\boldu} \! +\! \tilde \boldu \! \cdot\!\! \nabla \tilde\boldu \!\! =\!\! \nu \nabla^2 \tilde\boldu  
   \!  - \!\!\nabla \! \cdot \!\! \boldT \!\!
      \! -\!\! (\nabla {{p}}_0\!\cdot\! \hat{e}_x\!\!+\!\!\nabla\! \tilde{p}^{\prime\!})\!/\!\rho_{\!f},% \noalign{\vskip{-5pt}}%[-10pt]%\nonumber
\end{eqnarray}
%\vspace*{-0.2cm}
%\begin{equation}
% \label{eq:lesChannelfluidmasseqn}
% \nabla\cdot \tilde{\boldu}=0,
% \end{equation}
% \vspace{-0.65cm}
%\begin{equation}
%\label{eq:lesChannelfluidmomeqn}
%%\frac{\partial \tilde{u}_i}{\partial t} + \frac{\partial \tilde{u}_i \tilde{u}_j}{\partial x_j} = \nu \frac{\partial^2 \tilde{u}_i }{{\partial x_j^2}} 
%%     - \frac{\partial T_{ij} }{{\partial x_j}}
%%      +f(t) \delta_{i1} - \frac{1}{\rho} \frac{\partial \tilde{p'} }{{\partial x_i}}, \\ % + F_i^{fs}+g\delta_i. +f(t) \delta_{i1}
%\partial_t \tilde{\boldu}  + \tilde \boldu\cdot \nabla \tilde\boldu  = \nu \nabla^2 \tilde\boldu  
%     - \nabla\cdot \boldT
%      +f(t) \hat{e}_x - \nabla \tilde{p'}/\rho_f, 
%\end{equation}
\vspace*{-.6cm}  \\ 
where  
$\tilde{\boldu}$ is the filtered fluid velocity, %time dependent   to enforce constant mass flux
$\nabla{p}_0\!\cdot\!\hat{e}_x$ is a source term corresponding to a fixed streamwise pressure gradient\,(necessary to maintain the flow because the atmospheric boundary layer is modeled as a half-channel), $\tilde{p}^\prime$ is the perturbed pressure,  $\nu$ is the kinematic viscosity and $\boldT \!\!=\! \!\widetilde{\boldu \boldu} \!-\! \tilde{\boldu}\tilde{\boldu}$ is the subgrid stress\,(SGS) tensor. 
The fluid equations are coupled with equations for the conservation of mass and momentum for charged solid sand particles. Considering the volume fraction of the sand phase is small\,($\!<\!\!10^{-6\!}$), we assume a one-way coupling coupling between the fluid and the solid phases. %, i.e., the fluid influences the solid phase and momentum is transferred from the fluid to the sand but not vice-versa.
 The  mass and momentum conservation for the solid phase is expressed  as,
\vspace*{-0.3cm}
 \begin{eqnarray}
\label{eq:partcmassindex}
&&\partial_t {\tilde{n}_s}  +  \nabla\cdot (\tilde{n}_s \tilde{\boldv}_s) = 0,\\[-4pt] %\nonumber \\
\label{eq:partcmomindex}
  &&{m_s}\partial_t \tilde{\boldv}_s + {m_s}\tilde{\boldv}_{s} \cdot \nabla \tilde{\boldv}
=  \tilde{\boldD}_s +  q_s {{\boldE}}, \\[-3pt]%\nonumber\\
 \label{eq:nondimparticleparameterselec} &&\tilde{\boldD}_s =  {3\pi \nu \rho_f } { d_s} \cdot  \left( 1+0.15Re_{s}^{0.687} \right)\cdot   (\tilde{\boldu}-\tilde{\boldv}_{s}),\nonumber\\[-5pt] %\vspace*{-.5cm}
 && {{\boldE}}=-\nabla\phi, \qquad \varepsilon \nabla^2 {\phi}  = \Sigma, \qquad \Sigma\!=\!\!\!\! \sum_{s=1}^{s=\mathcal{S}}\!\!\! \tilde{n}_{{s}}  {q_s}.%\nonumber 
 \end{eqnarray}
 \vspace*{-.5cm}  \\ 
The distribution function for the sand particles is sampled at $\mathcal{S}$ points, and hence ${\mathcal{S}(\!=\!2)}$ is the total number of species considered.
$\!{\varepsilon}$ is the dimensional permittivity of the atmosphere, and 
 $\Sigma$ is the local net charge density.
$\tilde{\boldv}_s$, ${\boldD}_s$, $q_s  {{\boldE}}$ and $\phi$ represent the velocity, drag force, electrostatic force of species $s$,  and the electrostatic potential field, respectively; and 
$Re_s\!\!=\!\! | {\tilde{\boldu}}-{\tilde{\boldv}_s}|{U}_{\infty} d_{s}/\nu$ is the Reynolds number of the particle. 
% It is to be noted that for the continuity equation of the solid phase, the discretization is based on an upwind scheme to better capture discontinuities in concentration. 
% In this work, we developed a positivity-preserving conditionally higher order upwind scheme, which can  maintain the positivity of particle number density~(Eq.~\ref{eq:partcmassindex}) at all %the next 
% time steps. %, provided that the parameter is positive at the previous time.
\begin{figure}[htb!]
\vspace*{-4.mm}
%\centering
%\begin{tabular}{c}
%\subfloat[]
%{\includegraphics[width=0.5\linewidth]{figures/computationalsetup.jpg}} & 
%\subfloat[]
%{\includegraphics[width=1\linewidth,height=.5\linewidth]{figures/Mech6.pdf}} % &
{\includegraphics[width=1\linewidth,height=.5\linewidth]{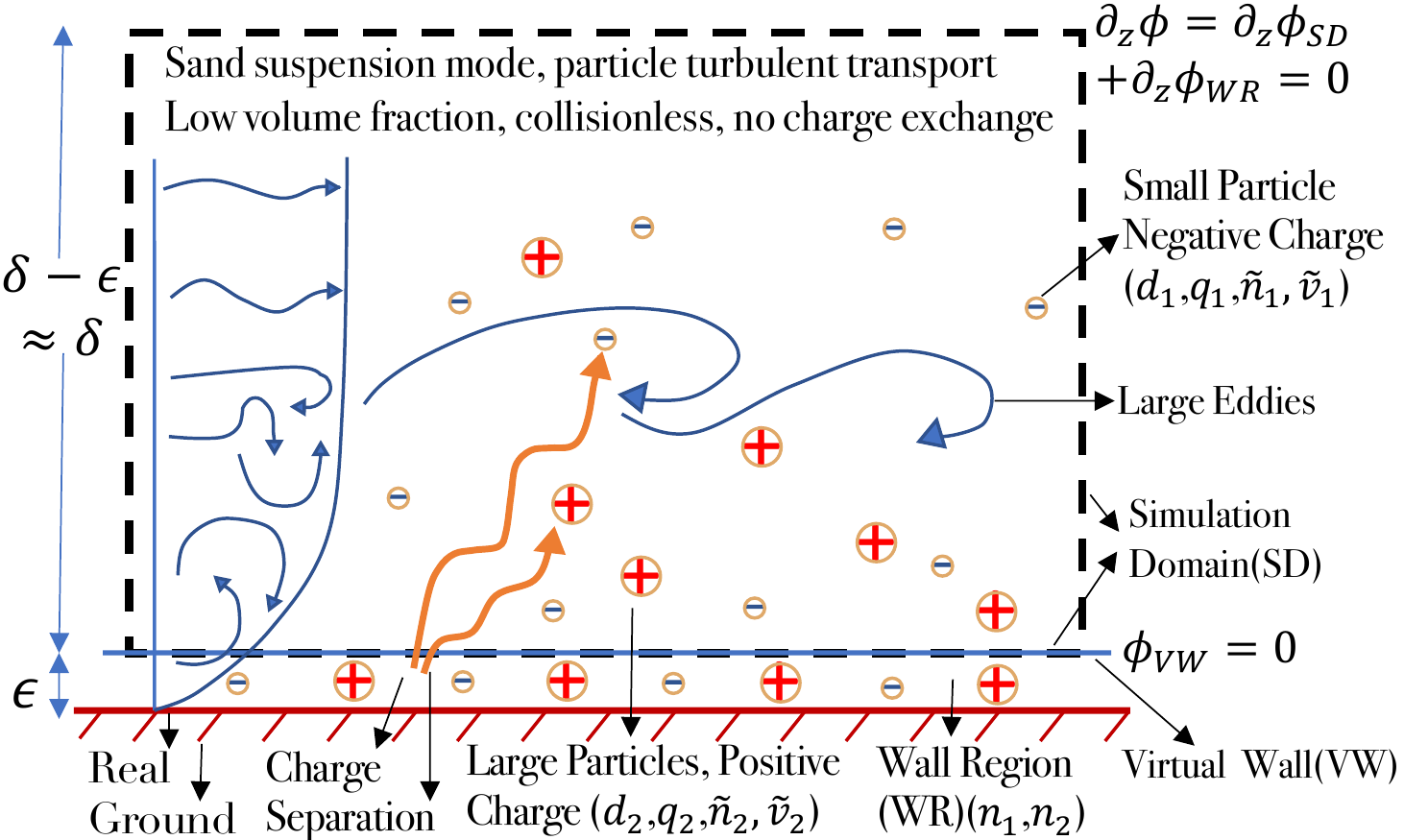}} % &
\vspace*{-8mm}
 \caption{Schematic of the simulation domain\,(SD) . The atmospheric boundary layer  height $\delta$ is modeled as a turbulent half-channel flow with streamwise and spanwise periodic conditions. The lower boundary of SD is modeled as a virtual wall\,(VW) at height $\epsilon\!\ll\! \delta$.  
 \vspace*{-5.mm}
}
  \label{fig:computationalsetup}
    \end{figure}
\begin{figure*}[htb!]
\vspace*{-5mm}
\centering
\begin{tabular}{cc}
%{\includegraphics[width=.50\linewidth]{figures/{Ez_time_xper.000385.000097.000002.dat_t_e_InterpAvg__ExpmZoom_xlim2}.pdf}} &
%{\includegraphics[width=.50\linewidth]{figures/{Ez_time_xper.000385.000097.000004.dat_t_e___ExpmZoom_xlim2}.pdf}} \\ 
{\includegraphics[width=.48\linewidth]{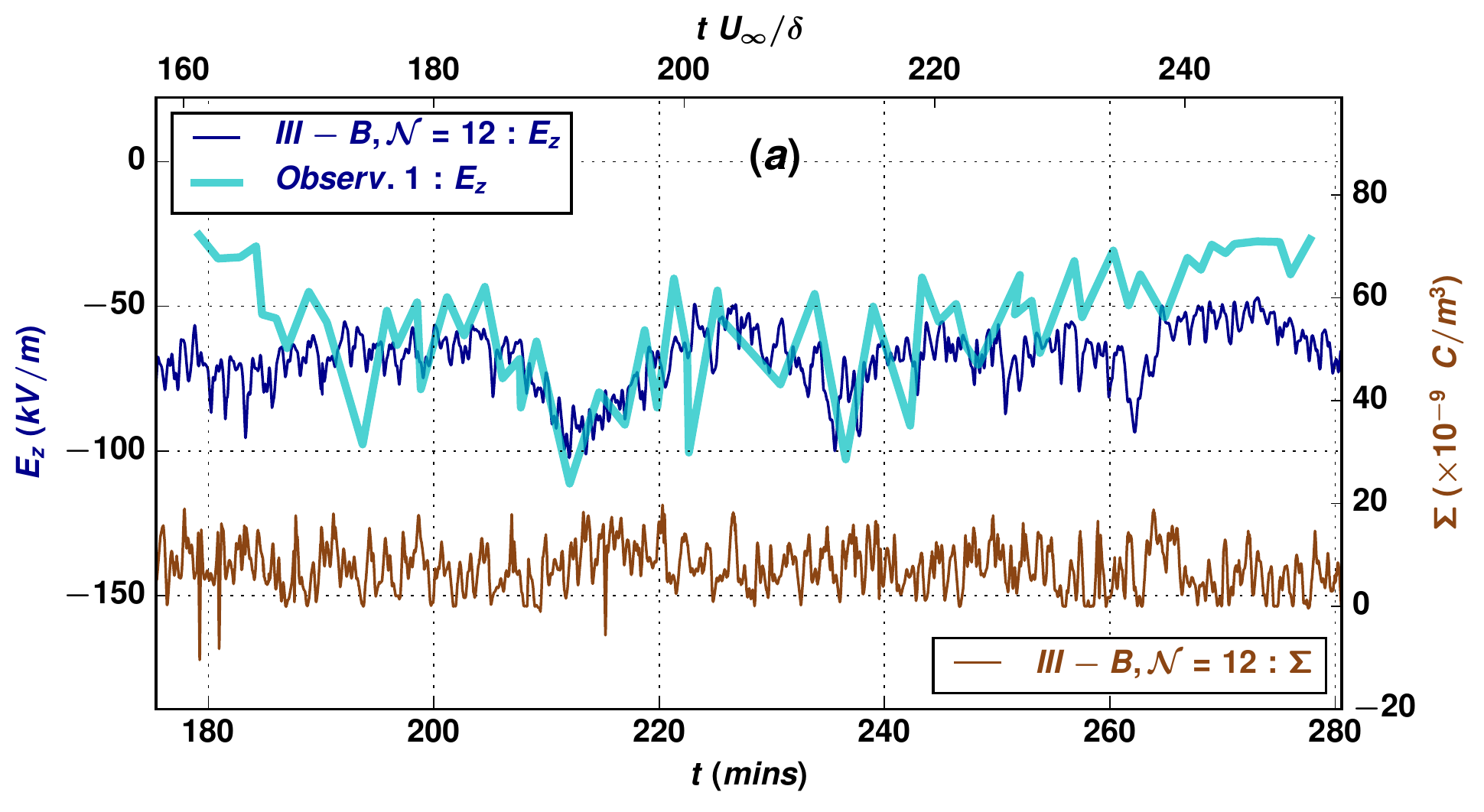}} &
{\includegraphics[width=.52\linewidth]{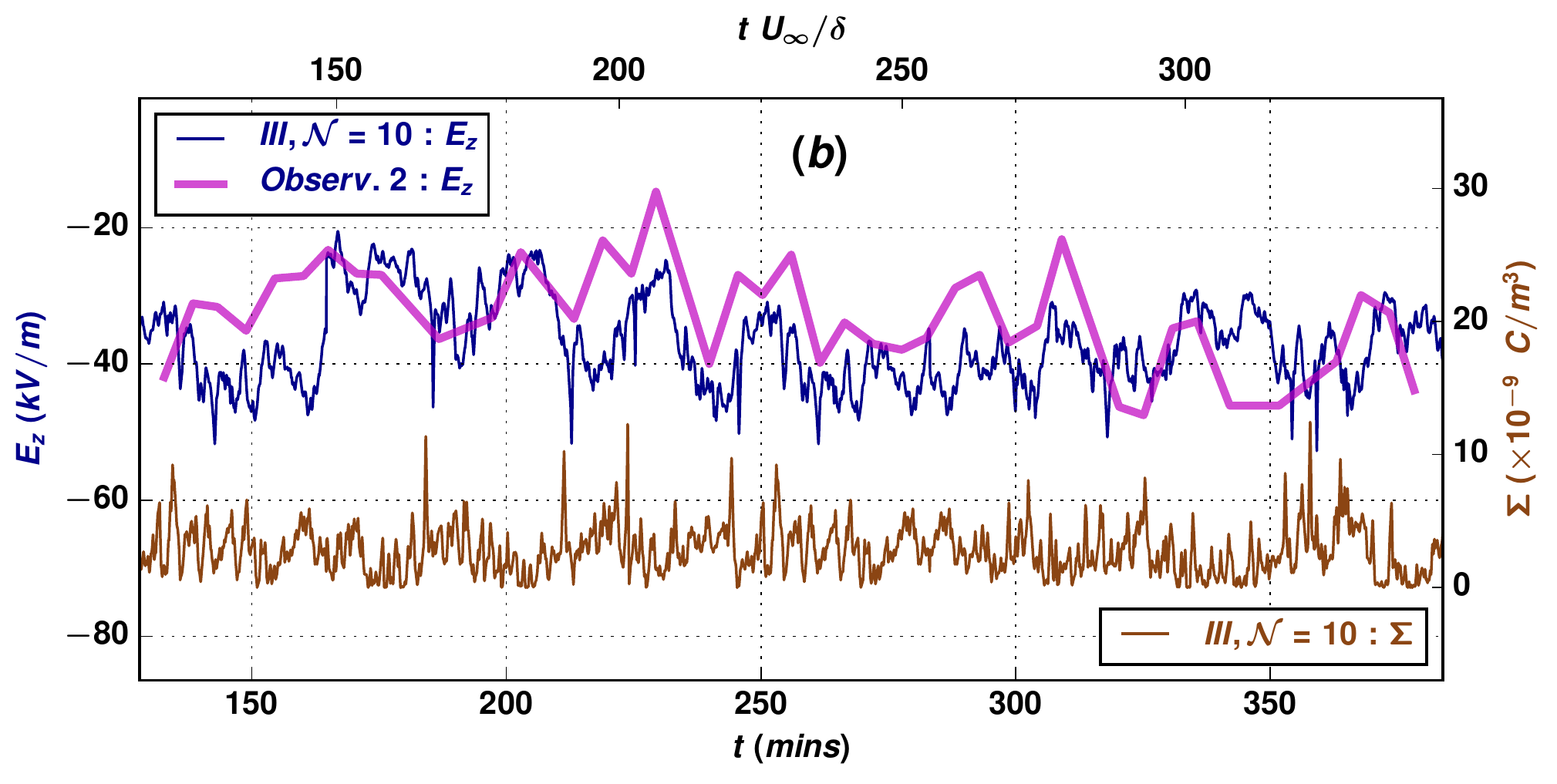}} \\ 
\end{tabular}
\onecolumngrid % and 
\vspace{-5mm}
 \caption{
 Time history of the simulated instantaneous wall-normal electric field\,(${E}_z$) at the mid-span mid-stream grid location\,(blue)  along with the charge density variation\,(brown) at (a)14.2$m$, Case III-B and (b)35$m$, Case III  where ${\mathcal{N}} =N_w/N_1$. Superimposed on the simulations are in-situ measurements  by Zhang {\it et al.}\,\cite{zhang2004experimental}\,(Observ. 1) and  Zhang {\it et al.}\,\cite{zhang2018variation}\,(Observ. 2). 
 }
\vspace{-7mm}
\label{fig:TimeVariationOfWallNormalEzAndChargeDensityVariation}
\twocolumngrid
\end{figure*}
%\section{Physical and computational setup}
\\{\it{Computational setup and boundary conditions.-}}
\label{sec:BoundaryConditions}
The above coupled set of equations governing the turbulent fluid flow with a suspension of charged sand particles is solved 
%in which
wherein 
the atmospheric boundary layer is modeled as a turbulent half-channel flow\,\cite{calaf2010large}.
The physical setup of the computational domain is depicted schematically in Fig.\,\ref{fig:computationalsetup}. 
The lower computational boundary is not at ground level but %a lifted-off distance
 at an elevated height\,(denoted as $\epsilon$) into the log-layer of the channel flow -- this is akin to the wall-modeled LES approach of Chung \& Pullin% for turbulent channel flow simulations
\,\cite{Chung2009281}. We use periodic boundary conditions in the streamwise and spanwise direction for both the fluid and solid phases. %At the top boundary, outflow conditions are imposed.
 At  the virtual wall\,(VW), the wall model leads to a dynamic Dirichlet boundary condition for the fluid velocity\,\cite{Chung2009281}. At the ground level, sand particles collide with each other, exchange charge and get entrained into the flow. The physics of such a %the
  charge exchange and lift-off process is complex. Here, we assume that the charge exchange process takes place below the virtual wall. One approach is to parameterize this phenomenon  as a flux boundary condition of charged sand particles into the flow, which is proportional to the number density, and dependent on many factors like soil humidity, ground temperature%, etc
  as well as other wind erosion factors, and not just atmospheric flow Reynolds number. 
An alternative, simpler, approach is to specify a Dirichlet boundary condition for the particle number density distribution\,($N_w$) at the virtual wall, i.e., an ${\mathcal{S}}$-vector of number density values is specified at the virtual wall. 
For the electrostatic potential we use a zero wall-normal gradient\,(Neumann BC) at the top% of the
\,($\partial_z\phi_{SD}\!\! =\! 0$), along with the  
Dirichlet %condition
BC\,(${\phi}_{VW}\!\!=\! 0$) 
at the virtual wall. 
%An alternative wall normal boundary conditions for the Poisson equation~(Eq.~\ref{eq:nondimparticleparameterselec} of electrostatic potential~($\phi$) 
%in Fig.~\ref{fig:computationalsetup}, can be given as zero wall normal-gradient~(Neumann BC) at the top of the domain~($\partial_z\phi_{SD}=0$), along with the  
%Dirichlet %condition
%BC~(${\phi}_{VW}=0$) 
%at the virtual wall. 
In order to maintain charge neutrality in the total domain\,(simulation domain\,(SD) plus the wall region\,(WR) % domain below the virtual wall
 $\!$), the excess charge density in the simulation\,(SD) is distributed equally but with an opposite sign in the region below the virtual wall. 
% Assuming typical wind speeds in the range $U_{\infty}=5-40$ m/s in a boundary layer of thickness $\delta\sim 100-3000$ m, the resulting Reynolds numbers is of order $Re_{\delta}=U_{\infty}\delta_o/\nu\sim 10^7-10^{10}$. Correspondingly, the Kolmogorov length is of order $\eta=\delta/ Re_{\delta}^{3/4}\sim 0.1-1$ mm.
% Sand particles have a  typical diameter of order $d_p\sim 0.01-1$ mm and a density $\rho_p\sim 2650$ kg/m$^{3}$. As a result, $d_p$ maybe much smaller than the smallest turbulent scales only when the Reynolds number is not too large, or the boundary layer is sufficiently thick.
 \\{\it{Numerical Methodology.-}}
\label{sec:Methodology}
The description of the gas phase follows an earlier work\,\cite{rahman2017modeling} on LES of incompressible turbulent flows\,(Eq.\,\ref{eq:lesChannelfluidmasseqn}). The simulations employ the stretched spiral-vortex SGS model along with a wall model that uses an inner-scaling ansatz to derive an ODE for a virtual-wall velocity. 
The numerical solver is based on a fractional-step method with an energy-conserving fourth-order finite-difference scheme on a staggered mesh\,\cite{rahman2017modeling}.
The dispersed solid phase\,(Eq.\,\ref{eq:partcmomindex},\ref{eq:partcmassindex}) is computed using the Eulerian approach of Direct Quadrature Method of 
Moments (DQMOM)\,\cite{desjardins2008quadrature}. 
% Important modifications are however required in the forces acting on the particles in comparison with regular treatments of electrically neutral multiphase flows.
% The aerodynamic drag forces on particles is modeled using a Stokes drag subject to a finite-Reynolds number correction. \cite{jiang2014analysis} %\cite{balachandar2010turbulent},
The dynamic electrical interactions between %within %between
charged particles are accounted for in the form of Gauss law\,(Eq.\,\ref{eq:nondimparticleparameterselec}),  %$F_{i,s}=q_s E_i$, where $\E_i$ 
which is solved using the multigrid technique. The solution to the Poisson equation governing the electrostatic potential takes into account the charge distribution below the virtual wall. %.
 The numerical code has been extensively tested and validated\,\cite{rahman2016large}.
\\{\it{Simulation Cases.-}}
\label{sec:ComputationalSetupCasesInvestigated}
For all the simulation cases reported here, we fix the free stream velocity of $U_{\infty}\!\!\!=\!\!\!15$m/s and a boundary layer of size $\delta\!\!=\!\!\!1000 m$\,(this corresponds to a Reynolds number based on boundary layer thickness to be $Re_{\delta}\!\!=\!\!10^{9}\!$, corresponding to a kinematic viscosity of air of $1.5\!\!\times\!\! 10^{-5\!} m^2\!/\!s$). 
 We sample the particle number density distribution function at two points\,($\!\mathcal{S}\!\!=\!\!2$), i.e., the sand phase is comprised of two sizes\,($d_1\!\!\!=\!\!200\mu m, d_2 \!\!\!=\!\! 500\mu m)$ of mass density $2650Kg\!/\!m^{3\!}$. 
 % (here chosen as as $d_{1}\!\!\!=\!\!\!250$ $\mu$m and $d_{2}\!\!=\!\!500$ $\mu$m. 
The representative\,(ideal) electric charge of the species of each size is  $Q_1\!\!\!\equiv \!\!(q_{1},q_{2})\!\!=\!\!(-4\!\times\!10^{-15}C ,2.24\!\times\!10^{-15}C)$.
A typical concentration\,\cite{liu2004experimental} %maximum amplitudes of the inflow number density 
at the bottom of suspension\,(top of saltation, O(1m)) are $N_1\!\!\!=\!\!(\!{n}_{1\!},\!{n}_{2\!})\!\! =\!\! (2\! \times\! 10^{7\!}m^{\!-3}\!,\!3.4\!\times\! 10^{6\!}m^{\!-3\!})$.
Four cases are considered in which the boundary conditions at the virtual wall $N_w$ is varied from $N_1$\,(corresponding to a weak storm) to $40 N_1$\,(corresponding to  a very strong sandstorm). These cases are labeled as: Case I\,($N_w\!\!\!=\!\!\!N_1$, ``weak"), 
Case II\,($N_w\!\!=\!\!4N_1$, ``moderate"), Case III\,($N_w\!\!\!=\!\!\!10N_1$, ``strong"),Case IV\,($N_w\!\!=\!\!40N_1$, ``very strong").
%The various simulation cases and other computational parameters are tabulated in Table~\ref{tab:LESChannelCases}.
 Case III-B\,($N_w\!\!=\!\!12N_1$, close to Case III) is included because it corresponds to field measurements of strong sandstorms. 
 Other parameters used in simulations are as follows. The fluid simulation domain is $x\!\!=\!\!32\delta$, $y\!\!=\!\!8\delta$, and $z\!\!=\!\!\delta$,  with $768$, $192$ and $96$ grid points in the $x$, $y$ and $z$ directions, respectively. The solid simulation domain in the wall-normal direction was truncated at $z\!\!=\!\!0.5 \delta$ to capture the near ground particle dynamics.
  The characteristic grid spacings in viscous wall units are $\Delta x^{+}\!\!\!=\!\!16\!\times\!10^5$, $\Delta y^{+}\!\!\!=\!\!16\!\times\!\!10^5$ and  $\Delta z^{+}\!\!\!=\!\xi^+\!\!\!=\!4\!\times\!\!10^5$. The virtual wall height $\epsilon$ in terms of viscous wall units is at $z^+\!\!\!=\!\!1.5\!\!\times\!\! 10^4$. 
\begin{figure*}[htb!]%[t!]
\vspace{-5.73mm}
%{\includegraphics[width=1.\linewidth]{figures/{0HeightVsMean_Ez_eVsSqrRtMSqrEz_eVsChargeDensity_e_16_9}.pdf}} \\ 
{\includegraphics[width=1.\linewidth]{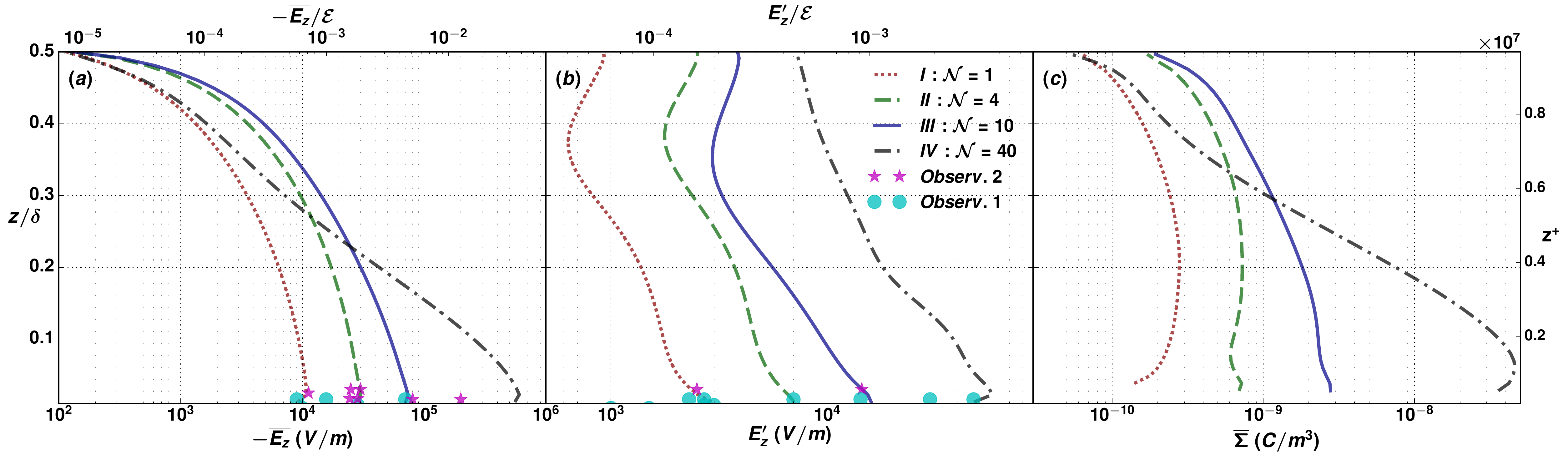}} \\ 
\onecolumngrid % and 
\vspace{-5.5mm}
\caption{  Altitude variation of the wall-normal electric field (a) mean (time and streamwise/spanwise averaged) $\bar{E}_z$, (b) RMS of fluctuations ${E}_z^\prime$, and (c) mean charge density, where $\mathcal{E} ={\delta \ {{\mathbf{\mathrm{M}}}(n_{_{1}}|q_{_{1}}|,n_{_{2}}|q_{_{2}}|)}}{/\varepsilon}$ 
is a reference electric field value corresponding to Case I ($M(a,b)=$mean of $a$ and $b$).
 %${\mathbf{\mathrm{M_h}}}(a,b)$ is the harmonic(${\mathbf{\mathrm{M_h}}}$) mean of $a$ and $b$. 
 The symbols  are mean and RMS values synthesized from various sandstorm measurements of Zhang {\it et al.}\,\cite{zhang2004experimental}\,(circles) and  Zhang {\it et al.}\,\cite{zhang2018variation}\,(stars).% $\eta = \alpha 
}
    \vspace{-7mm}
\label{fig:AltitudevariationofmeanRMSfluctuationsofwallnormalelectricfieldandmeanChargeDensity}
\twocolumngrid
\end{figure*}
\\  {\it{Results.-}}
Before we present results of the turbulence simulations, we remark that  we performed simulations for the fluid laminar regime comprising of a Poiseuille velocity profile. These laminar simulations yielded RMS electric field values smaller by orders of magnitude than those from turbulent simulations, % leading further credence
 further lending  credence
to our hypothesis. %that the large scale electric fields are sustained due to turbulent transport of charged sand particles. 
 \\{\it{Comparison with Field Observations.-}}
 %We compute the flow until it is statistically steady following which the vertical component of the electric field $E_z$ is time averaged over the spanwise domain. 
 The time variation of ${E}_z$ is plotted in 
 Fig.\,\ref{fig:TimeVariationOfWallNormalEzAndChargeDensityVariation} wherein we superimpose our simulation results with field observations\,\cite{zhang2004experimental,zhang2018variation}.  
 %The case III matches with the 
 %by Zhang {\it et al.} \cite{zhang2004experimental}.
 Since the time origin is somewhat arbitrary we align the minimum of ${E}_z$\,(Fig.\,\ref{fig:TimeVariationOfWallNormalEzAndChargeDensityVariation}(a)) and the pattern variation of ${E}_z$\,(Fig.\,\ref{fig:TimeVariationOfWallNormalEzAndChargeDensityVariation}(b)) between the observations and simulations. The overall average magnitude of ${E}_z$ from simulations compares well with the field observations although it is evident that the simulations also depict higher frequencies in the variation of ${E}_z$ compared with the observations\,(which may be limited due to instrumentation). %depicting such flow configuration can capture the  electric fields comparable to the field observations. 
\\  {\it{Varying sandstorm strength.-}}
 Cases I--IV are chosen to progressively increase in strength from weak to very strong sandstorms. 
 We compute the flow until it is statistically steady following which the vertical component of the electric field ${E}_z$ is time-averaged over the horizontal plane %. 
 %We compute the time and spanwise average of the wall-normal component of the electric field
  $\overline{{E}}_z$ and the RMS
  \,(root-mean-square) of ${E}_z^\prime$. For all four cases, the altitude variation of $\overline{{E}}_z$ and  ${E}_z^\prime$, and the mean charge density $\overline\Sigma$ are plotted in Fig.\,\ref{fig:AltitudevariationofmeanRMSfluctuationsofwallnormalelectricfieldandmeanChargeDensity}.
% Fig.~\ref{fig:AltitudevariationofmeanRMSfluctuationsofwallnormalelectricfieldandmeanChargeDensity} plots the altitude variation of wall-normal electric field and charge density for various particle parameter cases of the bottom boundary.
 The mean net electric field  Fig.\,\ref{fig:AltitudeVariationOfScalednetelectricfieldmagnitudesformeanRMSfluctuationsandratios}\,(a) of the cases decreases with height because of the decrease in the mean profile of charge density. 
 As the boundary number density\,($N_w$) increases, the mean and fluctuation magnitudes of electric field also increases. For each case, the magnitude of mean and RMS values decrease with altitude. 
The mean magnitude of the horizontal components of the electric field is negligible because of streamwise and spanwise periodic boundary conditions, but the %e 
RMS values are of the same order of magnitude as the vertical component.  
For the sandstorm case III, we note %with
the instantaneous maximum ${|{\boldE}|\!\!=\!\! 200kV\!/\!m}$ and range of 
%${\bm{E}}_z\!\!\!\in \!\!\!(-70,\!130)kV\!\!/\!m$ 
${{E}}_z\in (-70,\!130)kV\!\!/\!m$ 
in the domain. For
cases III, III-B and IV, the maximum magnitude of horizontal
electric field exceeds 100$kV\!/\!m$, which is
also observed in the field\,\cite{jackson2006electrostatic}.
%The mean values of case III\,($ \overline{E}_z\!\! =\!\!\!-75kV\!/\!m\!$) match field observations by Zhang {\it et al.}\,\cite{zhang2004experimental}.
For the weak sandstorm case\,(Case I) the near wall average electric field\,($\overline{{E}}_z$) are close to the observed electric field,\,($\overline{{E}}_z\!\!\approx\!\!-10KV\!/m)$\,\cite{zhang2018variation}, while for
Case II\,(moderate)\,$\overline{{E}}_z\!\!\approx\!\!-30KV\!/m$ and Case III\,(strong sandstorm)\,observations $\overline{{E}}_z\!\!\approx\!\!-80KV/m$, agree with field measurements\,\cite{zhang2004experimental,zhang2018variation}. %by Zhang {\it et al.}\,\cite{zhang2018variation}.
The RMS fluctuation of the near  wall vertical electric field  for the case II, Fig.\,\ref{fig:AltitudevariationofmeanRMSfluctuationsofwallnormalelectricfieldandmeanChargeDensity}(b),
is close to the $7kV\!\!/\!m$ observed in the field\,\cite{bo2013field}. %\cite{yair2016electrified}.  
 %For various cases, as the bottom wall charge density (corresponding number density) boundary condition for each species  increases, the magnitude of the electric field and net charge concentration increases. 
 The instantaneous electric fields % produced in sandstorm can be
 are in the same or opposite direction of the Earth's background electric field.
The instantaneous fluctuations at some locations are sufficiently high to cause a reversal in the direction of the vertical electric field component. %reverse sign. %can be much larger and these may
Such a change in the direction of the electric field has been observed in the field\,\cite{zhang2018variation}, and similarly  observed in the saltation layer\,\cite{schmidt1998electrostatic}. 
An incremental change in particle concentration boundary conditions tends to increase the
electric field  levels in the sandstorms.
%It is observed from Fig.~\ref{fig:AltitudevariationofmeanRMSfluctuationsofwallnormalelectricfieldandmeanChargeDensity}~(a) \& (b) that the ratio
%of fluctuations to means of the vertical electric field can be more than one.
%, but
%the ratios of the net electric fields are less than one near the wall,
%  Fig.~\ref{fig:AltitudeVariationOfScalednetelectricfieldmagnitudesformeanRMSfluctuationsandratios}.
%These vertical direction ratios are high because of the streamwise coupling.
The mean charge density plotted in Fig.\,\ref{fig:AltitudevariationofmeanRMSfluctuationsofwallnormalelectricfieldandmeanChargeDensity}\,(c) shows that its magnitude is at its maximum close to the wall. % and at the top of the domain.
 Although not shown, we note that the largest charge density fluctuation  also occurs in the vicinity of the wall and is well-correlated with the turbulent intensity of the flow.  %Computations of mean and RMS vertical electric field variation of the various cases near the ground in suspension mode displays that estimates of this framework stand in correspondence with experimental quantifications and field observations\cite{zhang2004experimental}, Fig.~\ref{fig:AltitudevariationofmeanRMSfluctuationsofwallnormalelectricfieldandmeanChargeDensity}.
\\{\it{Self-similarity.-}}
\label{sec:SelfSimilarity}
In the simulations considered above, the fluid turbulence is identical in all cases because of the one-way coupling between the fluid and the solid phases. The differences in the charge density stem from the differences in the flux of the sand particles into the turbulent flow. 
%The non-dimensionalizing term is evident from the Eq.~\ref{eq:nondimparticleparameterselec} and 
%Consider $\xi$ % = ( {\delta \ {{\mathbf{\mathrm{M}}}(n_{_{w1}}|q_{_{e1}}|,n_{_{w2}}|q_{_{e2}}|)} } {/ \epsilon})$
% as the maximum electric field~
%%$E_{max} = {  %\frac{ 
% possible by considering infinite planes of uniform surface charge density
%distribution $ \left( {\delta \ {{\mathbf{\mathrm{M}}}(n_{_{w1}}|q_{_{e1}}|,n_{_{w2}}|q_{_{e2}}|)} } \right) $
%of boundary layer thickness~($\delta$). 
%If we consider
The fundamental electrostatic charge interaction scales as 
%Coulomb's law
$E \!\! \propto \!\! q/r^2$. % and $\phi \propto q^2/r$. 
Combining this with the fact that in dilute particle flow, the average inter-particle distance\,($\!\propto\!\! {\tilde{n}_s}^{\!-1\!/3\!}$) for a low volume fraction of particles leads to the the $2\!/\!3$ scaling power law for the electric field, i.e., 
 $E \!\!\propto\!\! 1\!/\!r^2 \!\!\propto\! {\mathcal{N}}^{2\!/\!3}$. %\propto(\alpha)^{2/3} 
%In this large scale turbulent particle flow, for a given particle concentration($r$ as a constant) the  %electrostatic %$\propto n^{-1/3}$
%electric field 
%energy at a location can be approximated to 
%charge on the particles as $E\propto q$.
%dissipation~($\epsilon$) of the kinetic energy of the large eddies ($E^2  \propto\epsilon \propto U^3/L $). 
%, but t
%The electrostatic potential energy of the entire particle system can be approximated to the kinetic energy of the large eddies ($U^2  \propto q^2 $). 
%turbulence kinetic energy wiki
%and  t
%Thus the electric field ($E\propto U^{3/2}\propto q^{3/2}$).
The compensated electric field magnitude (normalized by ${\mathcal{N}}^{2\!/\!3\!}$) of both the mean and RMS  as a function of normalized distance are plotted in  Fig.\,\ref{fig:AltitudeVariationOfScalednetelectricfieldmagnitudesformeanRMSfluctuationsandratios}. 
\\ {\noindent{\it{Summary.- }}}
%The Discussion should be succinct and must not contain subheadings.
We have presented a framework to simulate a sandstorm modeled as the flow of charged sand particles in a turbulent flow of a statistically steady atmospheric boundary layer. % modeled as the flow of charged sand particles in a turbulent flow field of a half-channel (modeling a statistically steady atmospheric boundary layer). The wall-modeled LES are conducted in a domain bounded by a virtual wall on the bottom. At this virtual wall, charged sand particles with varying strengths are influxed into the turbulent flow field. Sand particles undergo charge exchange in a very small region next to the ground while modeled as collisionless in the simulation domain.  The high Reynolds number turbulent flow field transports the small and large sand particles differently because of the difference in the Stokes number\,(response time), inertia, and electric charge on the sand particles. 
%Consequently, the turbulent flow results in a large scale separation of the charged particles, with large variations in local charge density. 
 %Under the electrostatic approximation,  the resulting electric field is quantified. 
%  The resulting %spanwise 
%  averaged and RMS values of the electric field agree well with field observations. 
%It is to be emphasized that this is a novel modeling exercise of simulating the suspension of particles in sandstorms.
%A collection of methods are described for the framework of simulating sandstorms and capturing the generated electric fields by the charged particles. 
%To the best of our knowledge, such intense time-resolved CFD calculations with turbulence effect have not been undertaken so far and are the first of its kind. 
The computed values match those observed in sandstorms. A further increase in the concentration of  electrically 
 charged particles can reach the breakdown field in air, and therefore can trigger lightning (not modeled here). 
%In this study, wall modeled LES of charged particle-laden turbulent flow has been performed with the goal of modeling electrical phenomena in sandstorms. A Eulerian-Eulerian framework is employed in which sand particle motion is characterized in terms of mean  and RMS fluctuation terms of the electric field. % particle number density and
Our analysis demonstrates that the charge and concentration of sand particles are crucially responsible for the dynamics in sandstorms.
The electric fields produced in sandstorm conditions can be in the same or opposite direction to Earth's normal electric field and decrease with altitude. 
 We propose a simple scaling $\overline{|{{\boldE}}|},\!|{\boldE}|^\prime \!\! \propto \!{\mathcal{N}}^{2\!/\!3}$ that holds well for weak-to-moderate strength sandstorms.
 We posit that the level and frequency of occurrence of atmospheric turbulence will increase\cite{storer2017global} in the coming decades with the impact of climate change, making such 
studies more relevant. 
%Another place where electrical discharges in dust storms has been postulated and detected is on Mars.
Even though this work concentrates on earthbound dust 
 suspension, our simulation framework can also be useful for modeling  severe Martian sandstorms with suitable parameters.
% While the present work concentrated on sandstorms, this approach is also suitable for snowstorms\cite{latham1964electrification}.
% Although the airborne larger sand grains seems to not much response to the turbulent flow compared to small grains, large scale electric fields appear in this suspension mode because of the occurrence of charge separation. 
% The simulations make use of a multigrid method in order to drastically
% reduce the computational cost otherwise incurred by the many-
% body electrostatic problem. This technique enables a fast com-
% putation of the Coulombic forces on the discrete particles while
% avoiding inherent inaccuracies arising in alternative, homo-
% genized formulations of the electrostatic problem.
\\ Authors acknowledge the support of KAUST under awards URF/1/1704-01-01 \& BAS/1/1/1349-01-1. 
Cray XC40 Shaheen II at KAUST was used for simulations.
%MMR and WC benefited from their visit to CTR during the
%2016 Summer Program and would like to thank Dr. Javier Urzay
% for useful discussions.
%The authors wish to thank Dr. Javier Urzay for 
%useful discussions and help in the project.
% \\
% \vspace*{-1.4cm}
% \\
% \begin{table*}[b]
%  \vspace*{-23mm} 
%  \begin{ruledtabular}
%  \begin{tabular}{c|c|c}
%   $\ \  Where$
%  \end{tabular} 
%  \end{ruledtabular}
% \end{table*}
{ \onecolumngrid
\vspace*{-5.8mm}
\twocolumngrid
}

\end{document}